\documentclass[12pt]{article}
\usepackage{sao1}
\usepackage{psfig}

\begin{document}

 \begin{center}

{\LARGE \bf New Probable Dwarf Galaxies in Northern Groups of the Local Supercluster}

\bigskip

{\large I. D. Karachentsev$^1$, V. E. Karachentseva$^2$,
   and W. K. Huchtmeier$^3$}

\bigskip

$^1$Special Astrophysical Observatory, Russian Academy of Sciences, Nizhnii Arkhyz, 369167 Karachai-Cherkessian Republic, Russia  \\
$^2$Astronomical Observatory, Kiev National University, ul. Observatorna 3, Kiev, 04053 Ukraine                                     \\
$^3$Max-Planck-Institut f\"{u}r Radioastronomie, Auf dem H\"{u}gel 69, D-53121 Bonn, Germany ╗ ╗                                        \\
\end{center}
Received February 1, 2007

\bigskip

{\large \bf Abstract}
We have searched for nearby dwarf galaxies in 27 northern groups with
characteristic distances 8--15 Mpc based on the Second Palomar Sky
Survey prints. In a total area of about 2000 square degrees, we have
found 90 low-surface-brightness objects, more than 60\% of which are
absent from known catalogs and lists. We have classified most of these
objects (~80\%) as irregular dwarf systems. The first 21-cm line
observations of the new objects with the 100-m Effelsberg radio
telescope showed that the typical linear
diameters (1--2 kpc), internal motions (30 km s$^{-1}$), and hydrogen
masses ($\sim2\times10^7M_{\odot}$) galaxies correspond to those expected for the dwarf
population of nearby groups.

\bigskip


\section{INTRODUCTION}
A clear definition of galaxy membership in a group is required to analyze
the kinematics and dynamics of the groups of galaxies. However, the groups
of galaxies are defined in different ways, depending on the adopted
selection algorithm and the galaxy sample to which this algorithm is
applied.

Supplementing the groups by adding new dwarf galaxies makes
it possible to study in more detail the galaxy luminosity function,
the morphological segregation in groups, and the dependence of the
chemical composition of dwarf galaxies and the star formation rate
in them on their immediate environment. The problem of discrepancy
between predictions of the Lambda-CDM-model (Klypin et al. 1999) and
observational data has not yet been solved: the currently known number
of dwarf galaxies in the Local Group is approximately an order of
magnitude smaller than the predicted number. Thus, finding new dwarf
galaxies in nearby groups remains a very topical problem.

The chance
of detecting and identifying dwarf galaxies decreases sharply with
distance. The groups selected from (either "optical," CGC (Zwicky et al.
(1961--1968), or "infrared," 2MASS (Cutrie and
Skrutskie 1998)) flux-limited catalogs contain virtually no dwarf galaxies.
At the same time, purposeful thorough searches for putative dwarf members
of the Local Volume in the POSS-II and ESO/SERC sky surveys out to
distances of 5--6 Mpc proved to be successful (for more detail,
see the Catalog of Nearby Galaxies (CNG) by Karachentsev et al.
(2004)). These searches doubled the number of confirmed dwarf
galaxies in the nearest groups, which made it possible to study
in detail the kinematics and dynamics of the Local Volume groups
(Karachentsev 2005). Based on the similarity between the functions
of the linear diameters of galaxies in the Local Group and the
nearest groups, we undertook independent searches for the dwarf
population in the more distant (10.4 Mpc) Leo-I group
(Karachentsev and Karachentseva 2004). This required decreasing
the limiting angular diameter of the sought-for dwarf members
of Leo-I from 0.6--0.5 arcmin to 0.4--0.3 arcmin. We found
36 candidates for dwarf objects in the Leo-I group. HI 21-cm line
observations with the 300-m Arecibo radio telescope confirmed the
membership of most of them in the group (Stierwalt et al. 2005).
This encouraging result allowed us to undertake searches for new
probable dwarf members in other nearby groups of the Local Supercluster.

As we see from Table 1, the number of detected candidates for new group
members depends weakly
on the distance to the group (in the interval under consideration)
and on its population. Note, however, that we found the largest number
of new objects in groups where the brightest member is an early-type
galaxy (NGC 1023, NGC 3607).

\section{RESULTS OF OUR SEARCHES}
 As the initial
sample, we took the 2004 version of the Catalog of Groups of Galaxies
of the Local Supercluster with
radial velocities $V_{LG} < 3100$ km s$^{-1}$ (Makarov and Karachentsev 2000).
We selected the groups that satisfied the following three conditions
from this catalog: (1) the number of galaxies in the group $N\geq 4$; (2)
the mean radial velocity reduced to the Local Group centroid
$<V_{LG}> = 550-1100$ km s$^{-1}$; and (3) a positive declination of the
group center. There were 43 such groups. Subsequently, we excluded
the groups located in the Virgo cluster from
our searches, because of the complex dynamical pattern and the difficulty
of clearly separating its subsystems. Table 1 gives a list of groups
studied. Its columns present the following characteristics:

--name of
the dominating galaxy in the group;

--mean radial velocity in km s$^{-1}$;

--number of group members with known radial velocities; the number of
elliptical galaxies among them is given in parentheses;

--morphological
type of the dominating galaxy on the RC3 scale (de Vaucouleurs et al.
1991);

--numbers of the POSS-II fields in which the searches were
constructed;

--number of detected presumed dwarf members of the group
without radial velocities; the number of new objects is given in
parentheses.

\begin{table}[t]
\caption{Groups in which dwarf galaxies were searched for}
\begin{tabular}{lrrrclc} \\ \hline

Group &    $<V_{LG}>$ & $N(N_E)$& Type        & & POSS-II field numbers&  Number of objects\\
  (1)     &  (2)  &  (3)         & (4)        & & (5)     &(6)           \\
\hline
NGC 1023  &  824  &    17 (1)    &      $-$3  & &   299; 355; 356     &    12 (12)          \\
NGC 2681  &  752  &      6 (2)   &      0     & &     211               &             0 (0)   \\
NGC 2841  &  653  &      8 (2)   &      3     & &     212               &             1 (0)   \\
NGC 3180  &  565  &      4 (0)   &      6     & &     316; 317          &          1 (1)      \\
NGC 3486  &  621  &      4 (0)   &      5     & &     438               &             0 (0)   \\
NGC 3507  &  957  &    18 (2)    &      3     & &    569; 570; 640      &      1 (0)          \\
NGC 3607  &  854  &    11 (7)    &      $-$2  & &    570              &          21 (17)    \\
NGC 3627  &  705  &      6 (0)   &      4     & &     641; 642; 713; 714&    4 (3)            \\
NGC 3686  &1038   &     7 (0)    &      4     & &    641; 642           &         1 (0)       \\
NGC 3726  &  926  &    20 (3)    &      5     & &    216; 217; 266      &      0 (0)          \\
NGC 3938  &  805  &      8 (2)   &      5     & &     266; 267          &          0 (0)      \\
NGC 3953  &1092   &   18 (4)     &      4     & &   170; 171; 216; 217  &  4 (2)              \\
NGC 3972  &  871  &      4 (0)   &      4     & &     171               &             0 (0)   \\
NGC 4062  &  728  &      4 (0)   &      5     & &     440; 379          &          1 (1)      \\
NGC 4088  &  826  &      8 (0)   &      5     & &     216; 217; 266; 277&    0 (0)            \\
NGC 4151  & 1051  &   10 (0)     &      2     & &   321; 322; 379; 380  &  0 (0)              \\
NGC 4183  &   972 &      4 (2)   &      6     & &     217; 218; 267     &       2 (1)         \\
NGC 4258  &   561 &      9 (0)   &      4     & &     216; 217; 267     &       5 (2)         \\
NGC 4274  &   957 &      8 (1)   &      2     & &     441               &           12 (3)    \\
NGC 4278  &   609 &    10 (3)    &      $-$5  & &    441; 442         &        6 (3)        \\
NGC 4346  &   793 &      5 (1)   &      $-$2  & &     218; 267; 268   &      1 (0)          \\
NGC 4490  &   602 &      9 (0)   &      7     & &     268; 322          &         3 (2)       \\
NGC 4559  &   740 &    13 (4)    &      6     & &    380; 381; 442      &     5 (2)           \\
NGC 5033  &   996 &    15 (1)    &      5     & &    323; 324; 382      &     3 (2)           \\
NGC 5194  &   611 &      7 (1)   &      5     & &     220; 269; 270; 323&   2 (0)             \\
NGC 5248  & 1087  &     6 (0)    &      4     & &    720                &           1 (1)     \\
NGC 5906  &   913 &    13 (4)    &      5     & &    176; 177           &        2 (2)        \\
\hline
\end{tabular}
\end{table}
We carried out our searches on the blue (B) and red (R)
POSS-II prints to a limiting angular size of 0.4--0.3 arcmin.
We examined wide neighborhoods of the groups in order that the
radius of the search region be twice the harmonic mean radius of
the group. The results of our searches are summarized in Table 2,
where the following data are presented:

--object name, where "d"
stands for "dwarf," followed by two four-digit sequences indicating
the
hours and minutes in right ascension and the degrees and minutes in
declination;

--right ascension and
declination for epoch J2000.0;

-- type of dwarf galaxy:Ir--irregular,
Sph--
spheroidal, dE--elliptical;

--major and minor axes measured on the blue print;

--name of the group to which the object presumably belongs;

--notes and identification with known lists made using the NED database.

All of the objects found have a low surface brightness. In the notes,
VLSB and ELSB stand for very low ($\sim25^m$ arcsec$^{-2}$) and extremely
low ($\sim26^m$ arcsec$^{-2}$) surface brightness, respectively.

Our
independent searches revealed objects that either were found previously
(Karachentseva and Karachentsev 1998 (KK); Karachetnsev et al. 2001
(KKH) or were included in the lists of other authors (Binggeli et al.
1990 (BST); Shombert et al. 1997 (D); Trentham and Tully 2002 (ComaI);
Cabanela 1999 (MAPS); Trentham et al. 2001 (TTV)). Table 2 includes
those of them that have no measured radial velocities. The object
d1223+2935, whose radial
velocity was determined with a large error, was left in Table 2 for further
observations. In addition, we included the object d1243+4127, a possible
member of the NGC 4736 group missed when the lists of presumed dwarf members
of the Local Volume were published, in the list. Several objects, to be
described in more detail below, stand out among those found.

(1) d0226+3325.
This is an object of extremely low surface brightness located 13 arcmin SW
of the galaxy NGC 925. This galaxy lies on the periphery of the NGC 1023
group and was included in the catalog of isolated galaxies (Karachentseva
1973) as CIG 105. HI mapping of the spiral NGC 925 and its neighborhood
revealed a hydrogen satellite connected with NGC 925 by a "bar" (Briggs
1980; Gottesman 1980). Pisano et al. (1998) pointed out that there is a
hydrogen cloud with a radial velocity of $V_h = 524$ km s$^{-1}$ and a line
FWHM $W_{50} \sim  40$ km s$^{-1}$ at 10 arcmin SW of NGC 925. Note
that the mean
measured velocity of NGC 925 is $V_h = 553$ km s$^{-1}$. The hydrogen mass
of the cloud was estimated to be $10^7M_{\odot}$ . The cloud is invisible on a
deep R-band image obtained by these authors with the WINM telescope.
The object we found is probably an extremely low surface brightness
object and is the optical counterpart of the "hydrogen cloud."

\begin{table}
\caption{List of new galaxies found in 27 nearby northern groups}
\begin{tabular}{lcccll}\\ \hline
      Name  &  RA (2000.0) DEC   &Type     & $(a\times b)$'       &   Group   &           Notes      \\
       (1) &       (2)            &  (3)      &     (4)         &    (5)     &              (6)          \\
\hline
d0224+4102 &   022420.7+410212  &  Ir    &    0.50x0.30   &     N1023  &                                  \\
d0226+3325 &    022652.8+332537 &  Sph   &  1.30x1.20     &      N1023 &   ELSB at 13' SW of925           \\
d0237+4136 &   023718.8+413607  &  Ir    &    0.40x0.35   &      N1023 &  LSB                             \\
d0238+4052 &    023851.2+405247 &   Ir   &     0.70x0.40  &     N1023  & VLSB                             \\
d0241+3653 &    024131.5+365327 &   Ir   &     0.70x0.40  &     N1023  & Wedge-shaped   \\
d0243+3759 &     024302.0+375926&   Sph  &   0.60x0.50    &     N1023  &                                  \\
d0245+3955 &    024530.8+395547 &   Ir   &     0.35x0.30  &     N1023  &                                  \\
d0245+3957 &    024550.7+395711 &   Ir   &     0.90x0.75  &     N1023  &                                  \\
d0246+3952 &    024600.6+395238 &   Ir   &     0.70x0.50  &     N1023  &                                  \\
d0246+3910 &    024612.4+391055 &   Ir   &     0.75x0.50  &     N1023  & Distant spiral ?                 \\
d0246+3249 &    024621.8+324945 &   Ir   &     0.65x0.35  &     N1023  & VLSB                             \\
d0246+3832 &    024649.0+383251 &   Ir   &     0.40x0.25  &     N1023  &                                  \\
d0921+5016 &    092157.1+501612 &  Sph   &  1.10x0.90     &     N2841  &  KKH 49                          \\
d1018+4109 &    101822.6+410958 &   Ir   &     0.60x0.40  &     N3180  & LSB, blueish         \\
d1106+1250 &    110610.5+125042 &   Ir   &     0.50x0.25  &     N3507  & D 640-9                          \\
d1110+1932 &   111037.6+193217  &  Ir    &    0.35x0.30   &    N3607   &                                  \\
d1112+1845 &   111257.5+184540  &  dE    &  0.70x0.40     &    N3607   &F 570-4, distant?          \\
d1114+1802 &   111422.9+180235  &  Ir    &    0.40x0.30   &    N3607   &Sph?                              \\
d1115+1758 &   111507.1+175815  &  Ir    &    0.25x0.25   &    N3607   &dE?                               \\
d1115+1755 &   111513.0+175545  & dE     &  0.30x0.25     &    N3607   &                                  \\
d1115+1756 &   111524.5+175635  &  Ir    &    0.50x0.40   &    N3607   &dE? 19.5$^m$(MAPS)                 \\
d1115+1801 &   111536.4+180108  &  Ir    &    0.45x0.40   &    N3607   &Sph? ELSB                         \\
d1115+1804 &   111548.1+180438  &  Ir    &    0.25x0.25   &    N3607   &VLSB                              \\
d1115+1756 &   111558.0+175620  &  Ir    &    0.40x0.30   &    N3607   &                                  \\
d1116+1757 &   111611.7+175700  &  Ir    &    0.35x0.30   &    N3607   &D570-5                            \\
d1116+1713 &   111621.0+171347  & Sph    & 0.25x0.25      &    N3607   &ELSB                              \\
d1117+1818  &   111702.0+181807 &   Ir   &     0.45x0.35  &     N3607  &                                  \\
d1117+1719  &   111708.1+171909 &   Ir   &     0.30x0.25  &     N3607  &                                  \\
d1117+1759  &  111722.7+175945  &  Ir    &    0.30x0.25   &    N3607   &VLSB                              \\
d1117+1815  &  111748.2+181500  &  Ir    &    0.60x0.50   &    N3607   &distant spiral?                 \\
d1117+1737  &   111756.9+173726 &  dE    &   0.70x0.30    &    N3607   &Sph?, 18.8$^m$(MAPS)               \\
d1119+1157  &   111914.6+115709 &   Ir   &     1.00x0.35  &     N3627  & 17.9$^m$ (MAPS)                   \\
d1119+1404  &   111921.5+140434 &   Ir   &     0.65x0.45  &     N3627  &                                  \\
d1119+1732  &   111921.9+173214 &   Ir   &     0.40x0.35  &     N3607  & Sph?                             \\
d1120+1332  &   112016.1+133249 &   Ir   &     0.60x0.35  &     N3627  & Satellite of N 3628  \\
d1121+1830  &  112153.8+183008  &  Ir    &    0.35x0.30   &     N3607  &                                  \\
d1123+1916  &  112341.8+191615  &  Ir    &    0.55x0.45   &     N3607  &                                  \\
d1123+1816  &   112355.0+181657 &   Ir   &     0.40x0.30  &     N3607  &                                  \\
d1124+1125  &   112410.9+112514 &   dE   &   0.40x0.30    &    N3627   &distant?                          \\
d1134+1709  &   113416.2+170946 &   Sm   &   0.70x0.40    &    N3686   &KK 107                            \\
\end{tabular}
\end{table}

\setcounter{table}{1}
\begin{table}
\caption{continued}
\begin{tabular}{lcccll}\\ \hline

d1142+5210  &   114230.1+521036 &   Ir?  &     0.30x0.25  &     N3953  &                                  \\
d1148+5555  &   114843.8+555545 &   Ir   &      0.60x0.45 &      N3953 & KK  110=TTV15, 16.6$^m$            \\
d1150+5546  &   115006.3+554657 &   Ir   &      1.00x0.70 &      N3953 &  KKH 73, 16.5$^m$                 \\
d1154+3635  &   115423.9+363504 &   Ir   &      0.70x0.60 &      N4062 &                                  \\
d1156+5548  &   115601.2+554846 &   Ir   &      0.45x0.30 &      N3953 &                                  \\
d1205+4342  &   120525.0+434227 &   Ir   &      0.80x0.50 &      N4183 & KK 121                           \\
d1212+4237  &   121218.0+423732 &   Ir   &      0.50x0.45 &      N4183 &                                  \\
d1214+2749  &   121442.3+274955 &   Ir   &      0.35x0.30 &     N4274  &               \\
d1214+2915  &   121443.4+291511 &   Ir?  &     0.30x0.30  &     N4274  &ComaI-19,Sph?, VLSB,              \\
d1215+2813  &    121541.2+281315&     Ir?&        0.40x0.30  &      N4274 &  distant spiral?\\
d1215+2917  &   121547.8+291728 &    Ir  &        0.30x0.30  &      N4274 &  ComaI-16, 19.5$^m$            \\
d1216+2928  &    121639.2+292846&     Ir &         0.30x0.25 &      N4274 &  ComaI-12, VLSB,            19.2R\\
d1217+4703   &    121710.1+470349&     Ir &         0.35x0.25 &      N4258 &  BTS 109, Sph?, VLSB,  18.5$^m$\\
d1217+2914   &   121747.2+291436 &   Ir   &        0.40x0.40  &      N4274 &  ComaI-9, 18.1$^m$             \\
d1217+2828   &    121748.6+282827&    dE  &       0.70x0.45   &      N4274 &  KDG95= KK 130, 19.6$^m$       \\
d1218+2838   &    121829.4+283845&    Ir  &        0.80x0.70  &      N4274 &  KDG98= KK131=BTS115, 16.5$^m$ \\
d1218+3003   &   121831.8+300336 &   Ir   &       0.70x0.35   &      N4274 &  18.2$^m$(MAPS)                \\
d1218+2938   &    121843.6+293804&    Ir  &        0.40x0.35  &      N4274 &  ComaI-17, ELSB, 19.9R        \\
d1218+2833   &    121857.2+283312&    dE  &      1.10x0.60    &      N4274 &  BTS116, 15.5$^m$              \\
d1219+4743   &    121906.5+474351&    Ir  &        0.60x0.50  &      N4258 &  KK 132, VLSB                 \\
d1219+4727   &    121933.8+472706&    dE  &      0.30x0.30    &      N4258 &  BTS118=KK134, 17.0$^m$        \\
d1219+4705   &    121936.8+470533&    Sph &      0.25x0.25    &      N4258 &  LSB                          \\
d1219+2939   &    121943.8+293933&    dE  &      0.60x0.40    &      N4274 &  BTS119, 17.5$^m$              \\
d1210+4700   &    122040.6+470003&    dE  &      0.30x0.30    &      N4258 &  BTS132=KK 136, 17.5$^m$, Ir?  \\
d1220+4649   &    122055.0+464945&    Sph &     0.30x0.25     &      N4258 &  ELSB                         \\
d1221+2929   &    122108.8+292926&    Ir  &       0.45x0.35   &      N4278 &  ComaI-13,VLSB, 19.6R         \\
d1221+2905   &   122145.8+290502 &   Ir?  &     0.35x0.30     &      N4278 &  distant spiral?             \\
d1223+2832   &    122309.5+283235&    Ir? &      0.30x0.30    &      N4278 &  VLSB                         \\
d1223+2935   &     122357.7+293547&   dE  &      0.70x0.65    &      N4278 &  Vh=765$\pm$57,  18.4$^m$(MAPS)    \\
d1224+4707   &    122412.0+470724&    dE  &     0.40x0.35     &      N4346 &  BTS134, 17.0$^m$              \\
d1228+4358   &    122844.9+435818&    Ir  &       4:   x 1:   &      N4490 &  VLSB                         \\
d1229+3056   &    122941.2+305641&    Ir  &       0.40x0.30   &      N4278 &  Sph?                         \\
d1230+3002   &    123025.8+300224&     Ir &        0.55x0.40  &      N4278 &  VLSB, KDG 146                \\
d1233+3806   &    123307.4+380658&    Ir  &       0.50x0.40   &      N4490 &  BTS142, 17.5$^m$             \\
d1238+3512   &    123854.6+351218&    Ir  &       0.40x0.35   &      N4559 &  VLSB                         \\
d1242+4115   &    124212.3+411509&    Sph &    0.60x0.60      &      N4490 &  VLSB                         \\
d1243+3228   &    124325.2+322855&    dE  &     1.10x0.60     &      N4559 &  BTS151, 16.0$^m$             \\
d1243+2956   &    124344.2+295603&    Ir  &       0.60x0.40   &      N4559 &  BTS152, 17.5$^m$           \\
d1243+3232    &    124345.3+323201&    Ir? &      0.35x0.30    &      N4559 &                               \\
d1243+4127    &    124355.7+412725&    Ir  &       1.40x0.60   &      N4736 &  missed in KK lists        \\
d1248+3158    &    124852.8+315815&    Ir  &       0.70x0.50   &      N4559 &  KK 165=BTS156, 17.0$^m$      \\
d1251+4704    &    125114.5+470406&    Ir  &       0.70x0.50   &        ---   & BTS157, 17.5$^m$        \\
d1310+3648    &    131058.7+364813&    Ir  &       0.35x0.25   &      N5033 &  VLSB                         \\
\end{tabular}
\end{table}

\setcounter{table}{1}
\begin{table}
\caption{continued}
\begin{tabular}{lcccll}\\ \hline

d1310+3649    &    131059.1+364943&    Ir  &       0.60x0.30   &      N5033 &  VLSB                         \\
d1311+3710    &    131106.6+371041&    Ir  &       0.90x0.65   &      N5033 &  KK 188, near N5005          \\
d1317+4423    &    131719.5+442348&    Ir  &       0.80x0.50   &      N5194 &  KK 194, VLSB                 \\
d1332+4949    &    133236.2+494949&    Ir  &       1.20x0.65   &      N5194 &  UGCA 361,dE?                 \\
d1337+0908    &    133733.2+090815&    Ir  &       0.25x0.25   &      N5248 &  VLSB                          \\
d1504+5538    &    150451.2+553842&    Ir  &       0.35x0.30   &      N5906 &                               \\
d1508+5515    &    150832.5+551548&    Ir  &       0.50x0.40   &      N5906 &  Knotty                    \\
\hline
\end{tabular}
\end{table}
(2) A
close "triplet" in the NGC 1023 group. Our observations (see Table 3)
show that the dwarf galaxies d0245+3955, d0245+3957, and d0246+3952
are members of a single group.

\begin{figure}[hbt]
\centerline{\psfig{figure=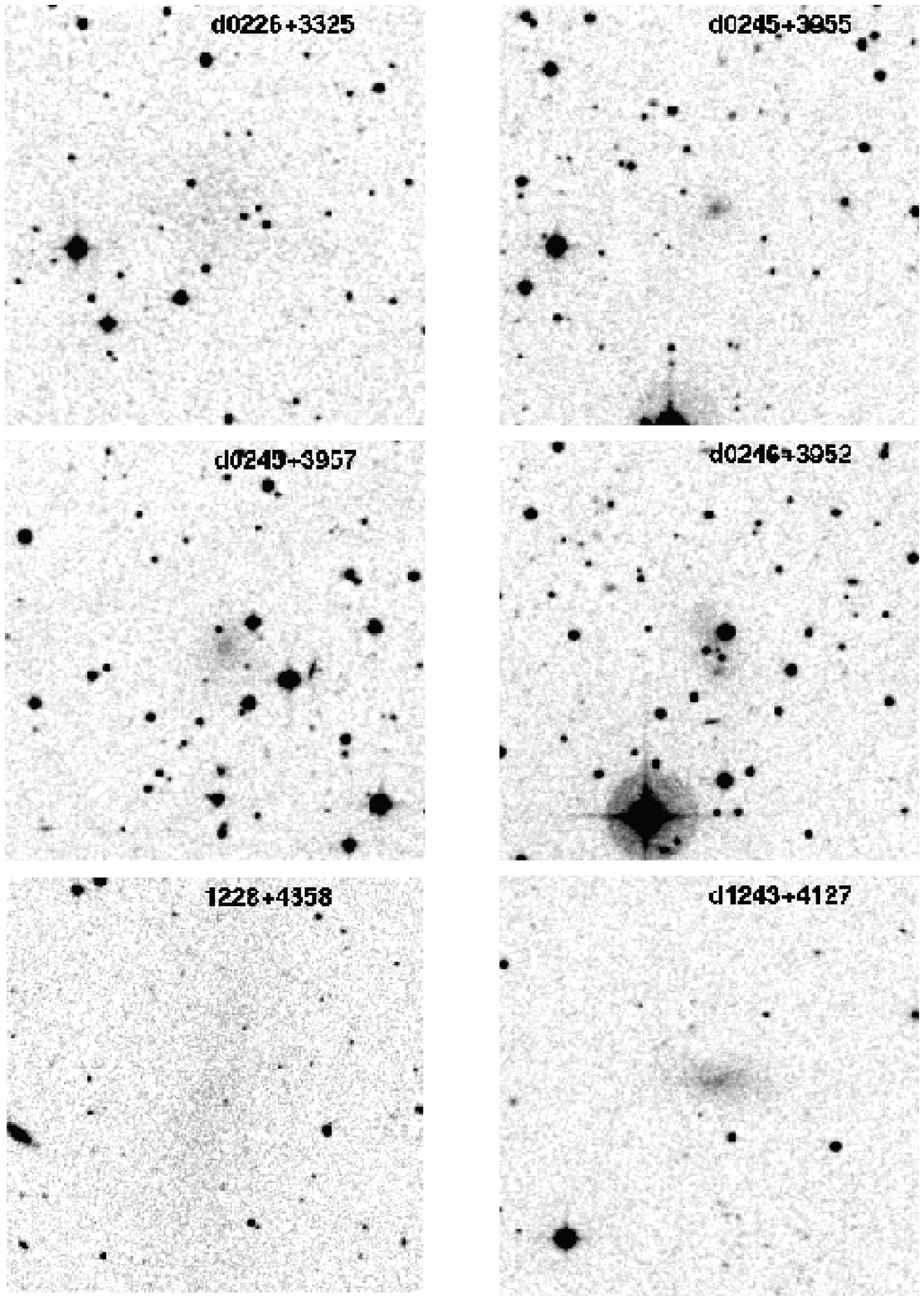,width=15cm}}
\caption{Reproductions of the images of six new low surface brightness dwarf galaxies
from the blue POSS-II prints. Each field is 5 arcmin$\times$5 arcmin in size;
north is at the top and east is on the left.}
\end{figure}

(3) d1228+4358. This object appears as
a partially destroyed "tail" pointing away from the galaxy NGC 4449
southward. It is similar in size, low surface brightness, and diffuse
structure to the dwarf system KK 208 with an old stellar population
near NGC 5236 (Karachentsev et al. 2002). This may be an example of
the formation of the so-called tidal dwarfs or, conversely, the capture
of a dwarf system by a bright galaxy just like the Sagittarius dSph
phenomenon around the Milky Way. The figure presents the images of
the six new dwarf galaxies mentioned above from the Digital Sky Survey.
Each image corresponds to the blue POSS-II prints. As we see from
Table 2, more than 60\% of the objects are absent from known catalogs
and lists. The median of the distribution of new objects in angular
diameter is 0.4 arcmin, which is half that for the known objects.
At a characteristic distance to the groups of 12 Mpc, the median
linear diameter of the new galaxies is about 1.5 kpc, which is typical
of the dwarf population of the Local Group and other nearby groups.
According to our estimates, about 80\% of the objects in Table 2 are
irregular dwarf galaxies. In such
systems, a significant fraction of the mass is usually accounted for by
neutral hydrogen (HI), which makes their 21-cm line observations promising.

\section{21-cm LINE OBSERVATIONS}

The objects from Table 2 were observed in 2005--2006 with the 100-m
Effelsberg radio telescope. At a beam width (FWHM ) of the radio
telescope equal to 9.3 arcmin, the angular sizes of the objects we
found occupy only a small fraction of its aperture. Therefore, a
considerable accumulation time is required to detect the HI flux
from them. Small angular separations between group galaxies are
another problem, which occasionally causes confusion of the signals
from several objects that fell within the aperture of the radio
telescope.

Table 3 presents the results of our observations for
several detected galaxies where no confusion arose because of
close neighbors. Its columns present the
following: 1, equatorial coordinates of the galaxy; 2, HI flux in Jy km s$^{-1}$;
3, maximum emission and (or) its rms error in mJy; 4, heliocentric radial
velocity and its error in km s$^{-1}$; 5, line full width at half maximum; 6,
total apparent magnitude of the galaxy that we estimated on the blue
POSS-II print compared to other galaxies of similar morphology; 7,
absolute magnitude of the galaxy corrected for the Galactic extinction
as derived by Schlegel et al. (1998) for an assumed distance to the
galaxy of $D = V_{LG}/H$, where $H = 72$ km s$^{-1}$Mpc$^{-1}$; 8,
HI mass-to-light
ratio in solar units, where the hydrogen mass was determined from
the HI flux $F$ as log $M_{HI} = \log F + 2 \log D + 5.37$. Judging by their
low radial velocities, weak HI fluxes, and small line widths, these
objects are actually dwarf galaxies with a typical hydrogen masses
of $\sim2\times10^7M_{\odot}$ . The HI survey of the galaxies listed in Table 2 is not yet
complete. We are going to publish the complete results of our observations
after the completion of the program.

\begin{table}
\caption{HI observations of several galaxies }
\begin{tabular}{clrclccc} \\ \hline
  RA (2000.0) DEC  &$F_{HI}$   &  Smax &  $V_{HI}$&  $W_{50}$  & $B_T$   & $M_B$ &$M_{HI}/L_B$        \\
(1)    &     (2)       & (3)     &   (4)   &   (5)   &        (6)  &(7)  & (8)\\   \hline
    024530.8+395547&    0.33 &     12$\pm$2  &    697$\pm$6      & 26 &    18.1&        $-$12.7 & 1.5\\
    024550.7+395711&    0.66 &     19$\pm$2  &    551$\pm$9      & 36 &    16.8&        $-$13.6 & 0.4\\
    024600.6+395238&    1.4  &        2  &    549$\pm$4   &    44 &    17.0&
	$-$13.4 & 1.0\\
    115006.3+554657&    0.5  &       8$\pm$2 &    594$\pm$6    &   20 &    16.5&        $-$13.4 & 0.3\\
    123307.4+380658&    0.46 &     21$\pm$3 &    719$\pm$4      & 23 &    17.5&
	$-$12.6 & 0.7\\
    124344.2+295603&    0.99 &     44$\pm$4 &   1141$\pm$3      & 22 &    17.5&
	$-$13.8 & 1.5\\
    124355.7+412725&    1.2  &      51$\pm$4&    402$\pm$2     &  16 &    16.2&
	$-$12.8 & 0.5\\
\end{tabular}
\end{table}

\section{CONCLUSIONS}
 Using the POSS-II prints,
we examined the regions of 27 northern groups of galaxies with expected
distances of about 12 Mpc (mean radial velocities in the range
550--1100 km s$^{-1}$) and with more than three members. The total
number of members in these groups with measured radial velocities
is 252. As a result of our searches, we added 90 more galaxies
without radial velocities, which we consider to be probable
members of these groups as low-surface brightness dwarf systems with
typical angular sizes of 0.4 arcmin, to this list. Most of them are
classified as dIr galaxies (80\%); the remaining galaxies are
classified as dE and dSph.

We observed some of the new objects
in the 21-cm HI line with the 100-m Effelsberg radio telescope.
These observations confirmed that the detected objects could be
attributed to the dwarf population of the groups under consideration
with a typical diameter of 1.5 kpc, an absolute magnitude of $-13^m$,
inner motions of 30 km s$^{-1}$, a hydrogen mass of $\sim2\times10^7M_{\odot}$, and an
HI mass-to-light ratio of $\sim1M_{\odot}/L_{\odot}$ . The radial velocity measurements
from the 21-cm line or optical spectra that we are planning will
allow us to improve the luminosity function of the galaxies in
nearby groups and to make the estimates of their virial masses
more reliable.

\bigskip

{\bf \large ACKNOWLEDGMENTS} This work was supported by
DFG-RFBR (grant no. 06--02--04017) and the Russian Foundation for Basic
Research (project no. 07--02--00005).

{\Large\bf REERENCES}

\bigskip
1. B. Binggeli, M. Tarengui, and A. Sandage, Astron. Astrophys. 228, 42 (1990).

2. E. Briggs, Astrophys. J. 238, 510 (1980).

3. J. E. Cabanela, PhD Thesis (Univ. of Minnesota, 1999).

4. R. M. Cutrie and M. F. Skrutskie, Bull. Am. Astron. Soc. 30, 1374 (1998).

5. S. T. Gottesman, Astron. J. 85, 824 (1980).

6. I. D. Karachentsev, Astron. J. 129, 178 (2005).

7. I. D. Karachentsev and V. E. Karachentseva, Astron. Zh. 81, 298 (2004) [Astron. Lett. 48, 267 (2004)].

8. I. D. Karachentsev, V. E. Karachentseva, and W. K. Huchtmeier, Astron. Astrophys. 366, 428 (2001).

9. I. D. Karachentsev, M. E. Sharina, A. E. Dolphin, et al., Astron. Astrophys. 385, 21 (2002).

10. I. D. Karachentsev, V. E. Karachentseva, W. K. Huchtmeier, and D. I. Makarov, Astron. J. 127, 2031 (2004).

11. V. E. Karachentseva, Soobshch. SAO RAN 8, 3 (1973).

12. V. E. Karachentseva and I. D. Karachentsev, Astron. Astrophys., Suppl. Ser. 127, 409 (1998).

13. A. Klypin, A. V. Kravtsov, O. Valenzuela, et al., Astrophys. J. 522, 82 (1999).

14. D. I. Makarov and I. D. Karachentsev, Astron. Soc. Pac. Conf. Ser. 209, 40 (2000).

15. D. J. Pisano, E. M. Wilcots, and B. G. Elmegreen, Astron. J. 115, 975 (1998).

16. D. J. Schlegel, D. P. Finkbeiner, and M. Davis, Astrophys. J. 500, 525 (1998).

17. J. M. Shombert, R. A. Pildis, and J. A. Eder, Astrophys. 111, 233 (1997).

18. S. Stierwalt, M. P. Haynes, R. Giovanelli, et al., Bull. Am. Astron. Soc. 37, 4 (2005).

19. N. Trentham and R. B. Tully, Mon. Not. R. Astron. Soc. 335, 712 (2002).

20. N. Trentham, R. B. Tully, and M. A. Verheijen, Mon. Not. R. Astron. Soc. 325, 385 (2001).

21. G. de Vaucouleurs, A. de Vaucouleurs, and H. C. Corwin, Third Reference Catalogue of Bright Galaxies (Springer-Verlag, New York, 1991), Vols. I╜III.

22. F. Zwicky, E. Herzog, M. Karpowich, et al., Catalogue of Galaxies and of Clusters of Galaxies (California Inst. Technology, Pasadena, 1961╜1968), Vols. I╜VI.
\end{document}